\documentclass[aps,prd,superscriptaddress,twoside,twocolumn,nofootinbib,10pt,%
showpacs,floatfix]{revtex4-1}
\usepackage{amsmath,amssymb}
\usepackage{graphicx,bm}
\usepackage{dcolumn}
\usepackage{epstopdf}
\usepackage{ulem} %% for strike-through
\usepackage[usenames]{color}
\usepackage{subfigure}
\usepackage{float}
\def\be{\begin{eqnarray}}
\def\ee{\end{eqnarray}}

\def\roughly#1{\mathrel{\raise.3ex\hbox{$#1$\kern-.75em%
\lower1ex\hbox{$\sim$}}}}
\def\lsim{\roughly<}

\def\la{\langle}
\def\ra{\rangle}

\def\bi{\bibitem}

\allowdisplaybreaks

\begin{document}

\title{Inhomogeneous quark condensate in compressed skyrmion matter}

\author{Masayasu Harada}
\email{harada@hken.phys.nagoya-u.ac.jp}
\affiliation{Department of Physics,  Nagoya University, Nagoya, 464-8602, Japan}

\author{Hyun Kyu Lee}
\email{hyunkyu@hanyang.ac.kr}
\affiliation{Department of Physics, Hanyang University, Seoul 133-791, Korea}

\author{Yong-Liang Ma}
\email{yongliangma@jlu.edu.cn}
\affiliation{Center of Theoretical Physics and College of Physics, Jilin University, Changchun,
130012, China}

\author{Mannque Rho}
\email{mannque.rho@cea.fr}
\affiliation{Department of Physics, Hanyang University, Seoul 133-791, Korea}
\affiliation{Institut de Physique Th\'eorique, CEA Saclay, 91191 Gif-sur-Yvette c\'edex, France}

\date{\today}
%%%%%%%%%%%%%%%%%%%%%%%%%%%%%%%
\begin{abstract}
The inhomogeneous quark condensate, responsible for dynamical chiral symmetry breaking in
cold nuclear matter, is studied by putting skyrmions onto the face-centered cubic crystal and
treating the skyrmion matter as nuclear matter. By varying the crystal size, we explore the
effect of density on the local structure of the quark-antiquark  condensate. By endowing the light
vector mesons $\rho$ and $\omega$ with hidden local symmetry and incorporating a scalar meson as a
dilaton of spontaneously broken scale symmetry, we uncover the intricate interplay of heavy mesons
in the local structure of the quark condensate in dense baryonic matter described in terms of skyrmion
crystal. It is found that the inhomogeneous quark density persists to as high a density as
$\sim 4$ times nuclear matter density. The difference  between the result from the present approach
and that from the chiral density wave ansatz is also discussed.
\end{abstract}
% 11.30.Qc Spontaneous and radiative symmetry breaking
% 11.30.Rd Chiral symmetries
% 12.39.Dc Skyrmions
% 12.39.Fe Chiral Lagrangians
\pacs{11.30.Qc,11.30.Rd,12.39.Dc}

\maketitle

%

%\tableofcontents

%\section{Introduction}

While the observation of the Higgs boson is considered to account for the masses of ``elementary
constituents" of visible matter, i.e., quarks and leptons, the bulk of the mass of the proton, the
constituents of which are  three light quarks, i.e., two up quarks and one down quark, remains more
or less unexplained. The quark masses account for only $\lsim 1\%$ of the proton mass. This is in
marked contrast to the next scale hadron, the nucleus. The mass of a nucleus of mass number $A$ is given
almost entirely, say, $\sim 98\%$, by the sum of the masses of $A$ nucleons.  It is generally
accepted, although not proven rigorously, that the nucleon mass arises from the nonperturbative
dynamics of strong interactions encoded in QCD. What is involved is quark confinement and chiral
symmetry spontaneously broken by the vacuum. It is not established but generally believed that the
two are related. It is easier to address the latter in effective theories, so we will focus on it in
what follows.

In QCD, the order parameter of the chiral symmetry breaking is the pion decay constant $f_\pi$ and,
in terms of the fundamental QCD quantities, the quark condensates, typically the two-quark condensate
$\langle \bar{q}q\rangle$. The two-quark condensate in cold or hot dense matter has been studied for a
long time. In Refs.~\cite{Campbell:1974qt,Campbell:1974qu} it was found that in neutron matter at
the neutron star density, both the $\sigma$ (the isoscalar component of the chiral four-vector) and
pion condensates exist and because of the chiral invariance of the system, these two condensates are
linked to each other through a chiral rotation. Later, the approach put forward in
Ref.~\cite{Campbell:1974qt} was investigated in more detail, and it was found that both condensates,
$\sigma$ and the pion, could depend on the position; that is, the condensates are
inhomogeneous~\cite{Dautry:1979bk}. In a phenomenological model with nucleons as explicit degrees of
freedom which are put on a crystal lattice, Pandharipande and Smith found that in solid neutron
matter, given an enhanced attractive tensor force, which is stronger than that given by standard
one-pion exchange, the inhomogeneous pion condensate could take place~\cite{Lattice}. Since then, the
inhomogeneity of these condensates,  now termed a chiral density wave(CDW), has been extensively
studied in various different effective  theory approaches modeling low-energy QCD; see, e.g.,
Refs.~\cite{Heinz:2013hza,CDW} and references therein.

In this article, continuing our effort to understand hadronic matter at high densities, we approach
this problem by putting skyrmions, representing nucleons, on a face-centered cubic (FCC) crystal
lattice. This has the advantage of enabling one to describe, on the same footing, both the basic
structure of the nucleon and the properties of nuclear matter, thereby probing the modification of
the vacuum in which the nucleons are propagating. By squeezing the crystal lattice, the density effect is
simulated and the properties of the condensates are modified. The model we employ is the
``generalized" skyrmion model that incorporates, in addition to the pion field carrying topology, the
vector mesons $\rho$ and $\omega$ introduced as hidden local symmetry fields and the dilaton $\chi$
associated with the broken scale symmetry of QCD. We call this the ``dHLS skyrmion" to distinguish it from the Skyrme model with pion field only. It turns out that each of these fields
plays an essential role in the structure of the nucleon as well as dense matter. In this article,  we
study the matter density effect and hadron resonance effect on the quark-antiquark condensates
at low and high densities of nuclear matter. The details of what we obtained before are involved and
given elsewhere (we refer the readers to Refs.~\cite{Ma:2013ooa,Ma:2013ela}). We summarize the key
points that are relevant to what we need for the discussions that follow.
\begin{enumerate}
\item There is a topology change from skyrmions to half-skyrmions at a density $n_{1/2}$ with or
    without nonpion fields. It might involve no symmetry change of QCD and hence no order
    parameter like what happens in condensed matter physics~\cite{science}. The location of
    $n_{1/2}$ depends, however, on the presence of the nonpion fields. Phenomenologically, it lies
    at $n_{1/2}
    \sim 2n_0$ where $n_0$ is the normal nuclear matter density.

\item The quark-antiquark condensate $\la\bar{q}q\ra\neq 0$ for $n<n_{1/2}$ as is appropriate for the
    Nambu-Goldstone phase. At $n_{1/2}$ the two-quark condensate, on average, in a unit cell
    vanishes, so $\overline{\la\bar{q}q\ra} =0$ for $n\geq n_{1/2}$ (here and in what follows,
    $\overline{\la\ra}$ stands for the space average in a unit crystal cell). However, the pion
    decay constant does not vanish, with massive hadrons excited, until a higher density
    $n_c>n_{1/2}$, with $n_c$ being the density at which the chiral symmetry is restored. What
    happens is that the two-quark condensate is not locally zero but vanishes, on average, in the
    unit cell. Thus chiral symmetry is not actually restored. This means that the two-quark
    condensate is not an order parameter for chiral symmetry in this model. There must be different
    order parameters for chiral symmetry, perhaps of multiquark nature, with the
    Gell-Mann-Oakes-Renner relation still holding in the phase with possible mutiquark condensates
    as an order parameter. The locally nonvanishing two-quark condensate immediately suggests the
    existence of an inhomogeneity in the half-skyrmion phase.

\item The nucleon mass can be characterized by two components as $m_N=m_0 +
    M(\overline{\la\bar{q}q\ra})$, where $M(\overline{\la\bar{q}q\ra})\rightarrow 0$ as
    $\overline{\la\bar{q}q\ra}\rightarrow 0$ and $m_0\neq 0$. Here, $m_0$ resembles the chiral-invariant mass term in the parity-doubled nucleon model~\cite{Heinz:2013hza}. Furthermore, as
    will be mentioned later, parity doubling does take place in the hadron spectrum in a cold or hot dense
    medium. However, in the dHLS model, the $m_0$ term which is related to the chiral doubling
    structure of nucleons does not figure explicitly in the Lagrangian, so it most likely reflects
    an emerging symmetry induced by the medium.

\item \label{here}There is an intricate interplay between the isovector vector meson $\rho$, the
    isoscalar vector meson $\omega$, and the scalar (dilaton) meson in dense baryonic matter.
    Suppose one generalizes the model used here by putting the infinite tower of isovector vector
    mesons $\rho$, $\rho^\prime$ etc. and $a_1$, $a_1^\prime$ etc. in the skyrmion model while
    dropping all isoscalar vector mesons and scalar mesons. Then one can arrive in flat space at
    many-nucleon systems - and infinite matter - that are described by BPS skyrmions with
    vanishing binding energies~\cite{sutcliffe}.  This is close to what is observed in experiments
    for binding energies of medium and heavy nuclei. What one notes is that each member of the
    tower of the isovector mesons tends to move the system closer to matter of BPS skyrmions. In
    the model used in this article, however, this tendency with the $\rho$ is spoiled by the
    isoscalar vector meson $\omega$.\footnote{The model with the infinite tower of isovector vector
    mesons arises in the large $N_c$ and large 't Hooft constant ($\lambda$) limit by dimensional
    deconstruction from 5D holographic dual QCD model that arises from string
    theory~\cite{sakai-sugimoto,HRYY}. With next-order $1/\lambda$ corrections, the isoscalar
    vector mesons coming from the Chern-Simons term and the warping of the metric, together,
    destroy the BPS structure, so one loses the nice agreement with nature. How this problem can be
    resolved is another matter discussed elsewhere~\cite{LR2015}.}

\end{enumerate}

In an effective model of QCD with linear chiral symmetry, the chiral four-vectors in the meson fields
$M$ are decomposed as
\begin{eqnarray}
M(x) = \sigma(x) + i \tau_a \pi_a(x), ~~\mbox{with $a = 1,2,3$},
\label{eq:decompM}
\end{eqnarray}
with $\tau^a$ as Pauli matrices. In the Nambu-Goldstone phase, the chiral four-vectors are constrained
by $\langle \sigma^2 + \pi_a^2 \rangle_{\rm QCD} = f_\pi^2$ (here and in what follows, $\langle
\rangle_{\rm QCD}$ stands for the medium-free QCD vacuum) in the QCD vacuum. Moreover, due to the
parity invariance of the QCD vacuum, one normally has $\langle \sigma \rangle_{\rm QCD} = f_\pi
\propto \langle \bar{q}q\rangle_{\rm QCD}$ and $\langle \pi_a \rangle_{\rm QCD} \propto \langle
\bar{q}i\tau_a \gamma_5 q\rangle_{\rm QCD} = 0 $. However, in a medium, due to the interaction between
the mesons and baryonic matter, both $\langle \sigma \rangle$ and $\langle \pi_a \rangle$ could be
nonzero and position
dependent~\cite{Campbell:1974qt,Campbell:1974qu,Dautry:1979bk,Lattice,Heinz:2013hza,CDW}.

In the nonlinear realization of chiral symmetry, the two-quark condensate is normalized as $\langle U
\rangle_{\rm QCD} = \langle M \rangle_{\rm QCD}/f_\pi = 1$ in the QCD vacuum with $U =
\exp(i\pi/f_\pi)$. However, in the nuclear matter medium, there can be deviation due to the
interaction between the pion and the matter medium, and the value of the deviation from $1$ accounts
for the deviation of the two-quark condensate from its vacuum value.

To explore the density dependence of the quark-antiquark condensate in a theory with the nonlinear
realization of chiral symmetry,   it is convenient to decompose the field $U(x)$ as in
\eqref{eq:decompM},
\begin{eqnarray}
U(x) = \phi_0(x) + i \tau_a \phi_a(x), ~~\mbox{with $a = 1,2,3$}.
\label{eq:decompU}
\end{eqnarray}
In the ground state of the system, $\phi_0(x)$ accounts for the isoscalar quark condensate while
$\phi_a(x)$ accounts for the isovector quark condensate which, in the QCD vacuum, have values
$\phi_0(x)= 1$ and $\phi_a(x) = 0$, but in the nuclear matter medium, they might have $\phi_0(x) \neq 1$
and $\phi_a(x) \neq 0$ due to the matter effect.
And, because of the cluster structure of nuclear matter, the quark condensates are expected to be
functions of space coordinates $\vec{x}$, i.e., inhomogeneous.

In the present work, to simulate the nuclear matter environment, among all the approaches to nuclear matter, we adopt the skyrmion matter approach and regard the skyrmion matter as nuclear
matter~\cite{Klebanov:1985qi,PV09}. Since in each crystal cell, the baryon number is fixed, the
density effect is simulated by changing the crystal size. Here, we adopt the FCC crystal~\cite{Kugler:1988mu,Kugler:1989uc} for which the nuclear matter density $n$ and crystal
size $L$ are related through the relation $n = 4/(2L)^3$. As stated at the beginning, in the skyrmion
crystal approach, one can treat both the nuclear matter and medium modified hadron properties in a
unified way~\cite{Lee:2003aq}.

To simulate nuclear matter from FCC crystal, it is convenient to define the unnormalized quantities
$\bar{\phi}_\alpha$~\cite{Kugler:1989uc,Lee:2003eg}, which are related to the quantities in the
decomposition \eqref{eq:decompU} through
\begin{eqnarray}
\phi_\alpha & = & \frac{\bar{\phi}_\alpha}{\sqrt{\sum_{\beta=0}^3 \left(\bar{\phi}_\beta\right)^2}},
\quad (\alpha,\beta = 0,1,2,3).\label{eq:norm}
\end{eqnarray}
In the crystal lattice, the unnormalized fields have the Fourier series expansions
\begin{eqnarray}
\bar{\phi}_0(x,y,z) & = & \sum_{a,b,c} \bar{\beta}_{abc} \cos(a\pi x/L)
\cos(b\pi y/L) \cos(c\pi z /L),
\label{sigma}\nonumber\\
\bar{\phi}_1(x,y,z) & = & \sum_{h,k,l} \bar{\alpha}_{hkl} \sin(h\pi x/L)
\cos(k\pi y/L) \cos(l\pi z/L),
\label{pi1} \nonumber\\
\bar{\phi}_2(x,y,z) & = & \sum_{h,k,l} \bar{\alpha}_{hkl} \cos(l\pi x/L)
\sin(h\pi y/L) \cos(k\pi z/L),
\label{pi2} \nonumber\\
\bar{\phi}_3(x,y,z) & = & \sum_{h,k,l} \bar{\alpha}_{hkl} \cos(k\pi x/L)
\cos(l\pi y/L) \sin(h\pi z/L).
\label{pi3}\nonumber
\end{eqnarray}
By varying the Fourier coefficients $\alpha$ and $\beta$, one obtains the minimal energy,
interpreted as the ground state of nuclear matter, for a specified crystal size $L$. The
inhomogeneous quark condensates with respect to their values in vacuum can then be calculated by
substituting the values of $\alpha$ and $\beta$ thus determined into Eq.~\eqref{eq:norm}. In addition,
by varying the crystal size $L$, the density effect on the quark condensate can be accessed. Note
that due to the FCC structure and the arrangement of the nearest-neighbor skyrmions to yield the
strongest attractive interaction, the modes appearing in the above equation are not
independent~\cite{Lee:2003aq}.

To explore the $\pi$, $\rho$, and $\omega$ meson effect on the inhomogeneous quark condensate, we use
the skyrmion model obtained from the hidden local symmetry approach given up to the next-to-leading
order in the chiral expansion including the homogeneous Wess-Zumino term. The degrees of freedom involved
are then the pion, the lowest-lying $\rho$ and  $\omega$ as employed in~\cite{Ma:2012zm,Ma:2013ooa}
to explore the structure of the single baryon as well as baryonic matter. Explicitly, we consider
the models HLS($\pi$), HLS($\pi,\rho$), and HLS($\pi,\rho,\omega$) which are explicitly defined in
Refs.~\cite{Ma:2012zm,Ma:2013ooa}. In addition to the hidden local symmetry degrees of freedom, to
explore the light scalar meson (i.e., dilaton) effect on nuclear matter, we apply the model
dHLS-II($\pi,\rho,\omega$) defined in Ref.~\cite{Ma:2013ela} in which the chiral restoration can be
realized at some high density $n_c$. The effective Lagrangian of model dHLS-II($\pi,\rho,\omega$) is
\begin{eqnarray}
{\cal L}_{\rm dHLSII} & = & {\cal L}_{\rm (2)}^{\rm dHLSII} + {\cal L}_{\rm (4)}^{\rm HLS}
+ {\cal L}_{\rm anom}^{\rm HLS} + {\cal L}_{\rm dilaton},
\label{eq:lagrdhlsII}
\end{eqnarray}
where ${\cal L}_{\rm (4)}^{\rm HLS}$
and ${\cal L}_{\rm anom}^{\rm HLS}$ are the $O(p^4)$ and homogeneous Wess-Zumino terms of the HLS Lagrangian, respectively, and
\begin{eqnarray}
{\cal L}_{\rm dilaton}  &=&  \frac{1}{2} \partial_\mu \chi \partial^\mu \chi \nonumber\\
& & {} - \frac{m_\chi^2 f_\chi^2}{4}\left[\left(\frac{\chi}{f_\chi}\right)^4 \left(\ln \left(\frac{\chi}{f_\chi}\right) - \frac{1}{4}\right) + \frac{1}{4}\right], \nonumber\\
\mathcal{L}_{\rm (2)}^{\rm dHLSII} & = &
f_\pi^2 \left(\frac{\chi}{f_\chi}\right)^2\mbox{Tr}[ \hat{\alpha}_{\perp\mu}
\hat{\alpha}_{\perp}^{\mu} ]
\nonumber\\
&&
{} + a f_\pi^2 \left(\frac{\chi}{f_\chi}\right)^2\mbox{Tr}[\hat{\alpha}_{\parallel\mu}\hat{\alpha}_{\parallel}^{\mu}]_{\rm SU(2)}
+ \frac{1}{2} af_\pi^2 g_\omega^2 \omega_\mu \omega^\mu
\nonumber\\
& & {} + \mathcal{L}_{\rm kin} , \label{eq:LagrdHLSII}
\end{eqnarray}
with the subindex ${\rm SU(2)}$ being the ${\rm SU(2)}$ subgroup of the hidden local symmetry ${\rm U(2)}$ and $\mathcal{L}_{\rm kin}$ being the kinetic term of the dilaton field. In the Lagrangian \eqref{eq:lagrdhlsII}, when the dilaton field is replaced by its vacuum expectation value, $\langle \chi \rangle = f_\chi$, the model dHLS-II($\pi,\rho,\omega$) reduces to HLS($\pi,\rho,\omega$). After switching off the homogeneous Wess-Zumino term, one arrives at HLS($\pi,\rho$) and, after integrating out the $\rho$ meson from HLS($\pi,\rho$) we obtain the model HLS($\pi$). For other details of the relevant models, we refer to Refs.~\cite{Ma:2012zm,Ma:2013ooa,Ma:2013ela}.

\begin{figure*}
\centering
\subfigure[~$L=2.5~$fm]{
\begin{minipage}[b]{0.21\textwidth}
\includegraphics[width=1\textwidth]{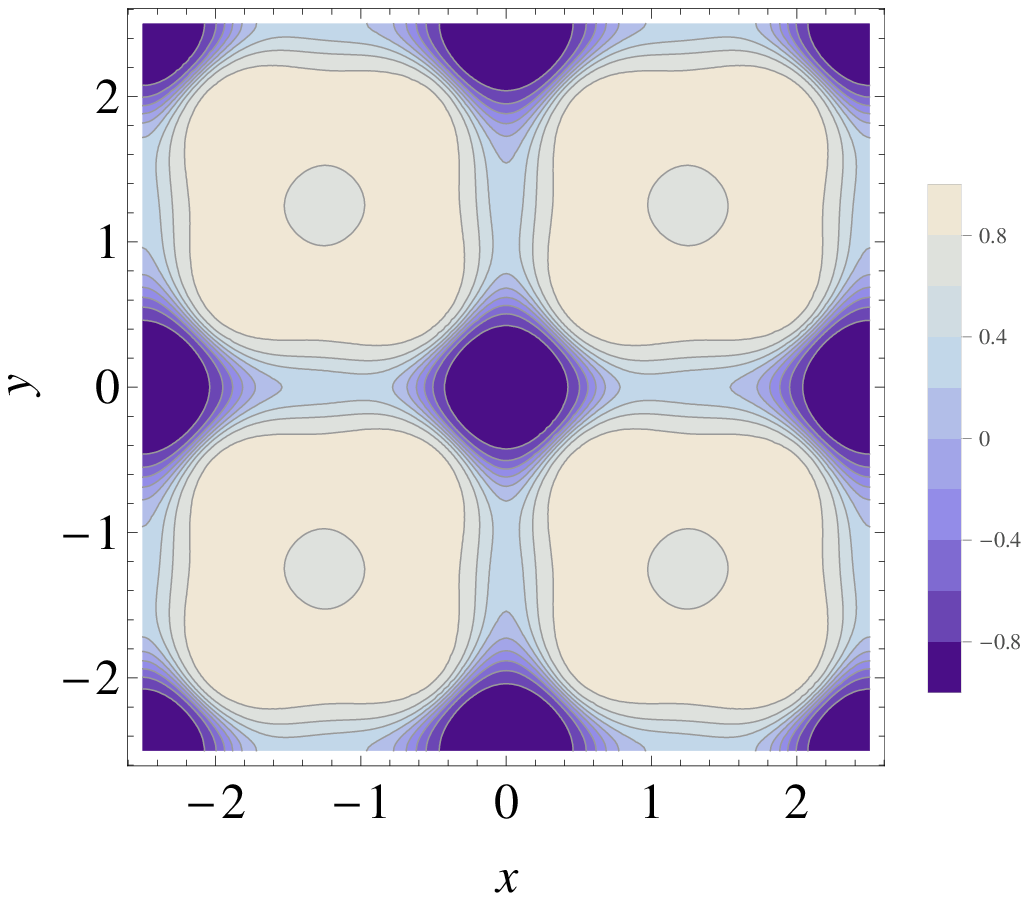} \\
\includegraphics[width=1\textwidth]{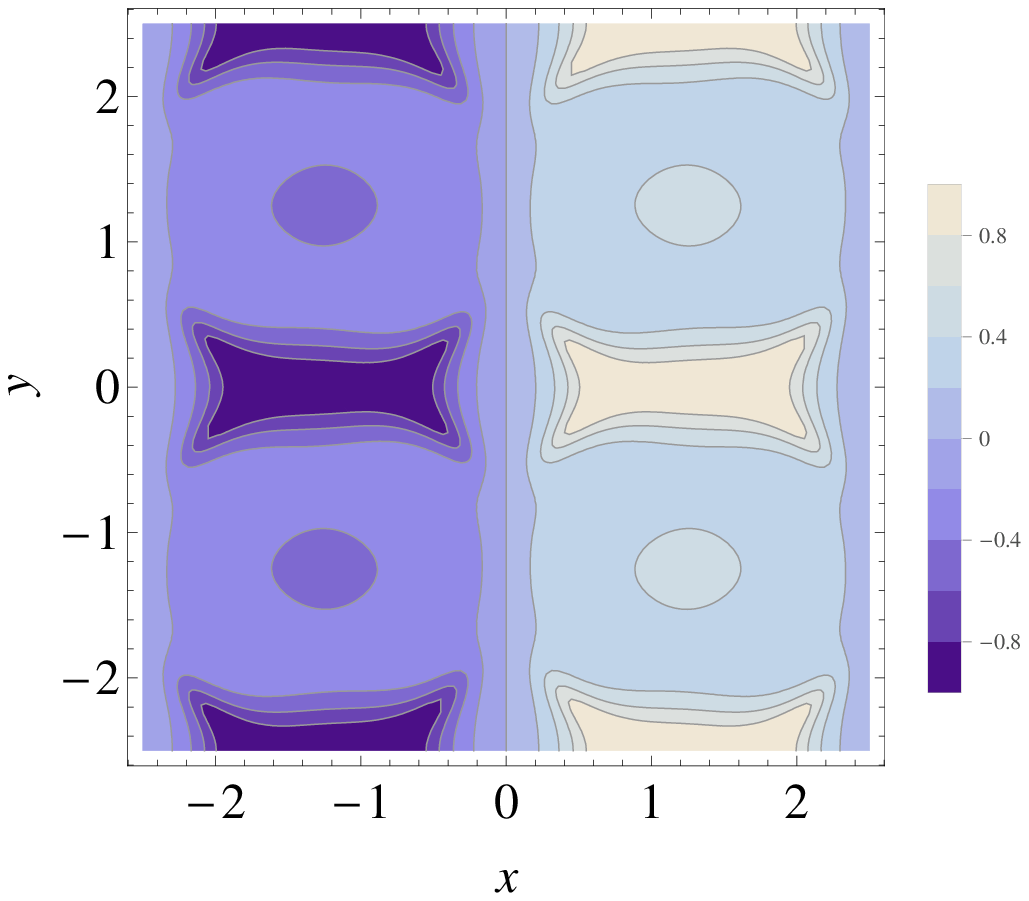}
\end{minipage}
}
\subfigure[~$L=1.5~$fm]{
\begin{minipage}[b]{0.22\textwidth}
\includegraphics[width=1\textwidth]{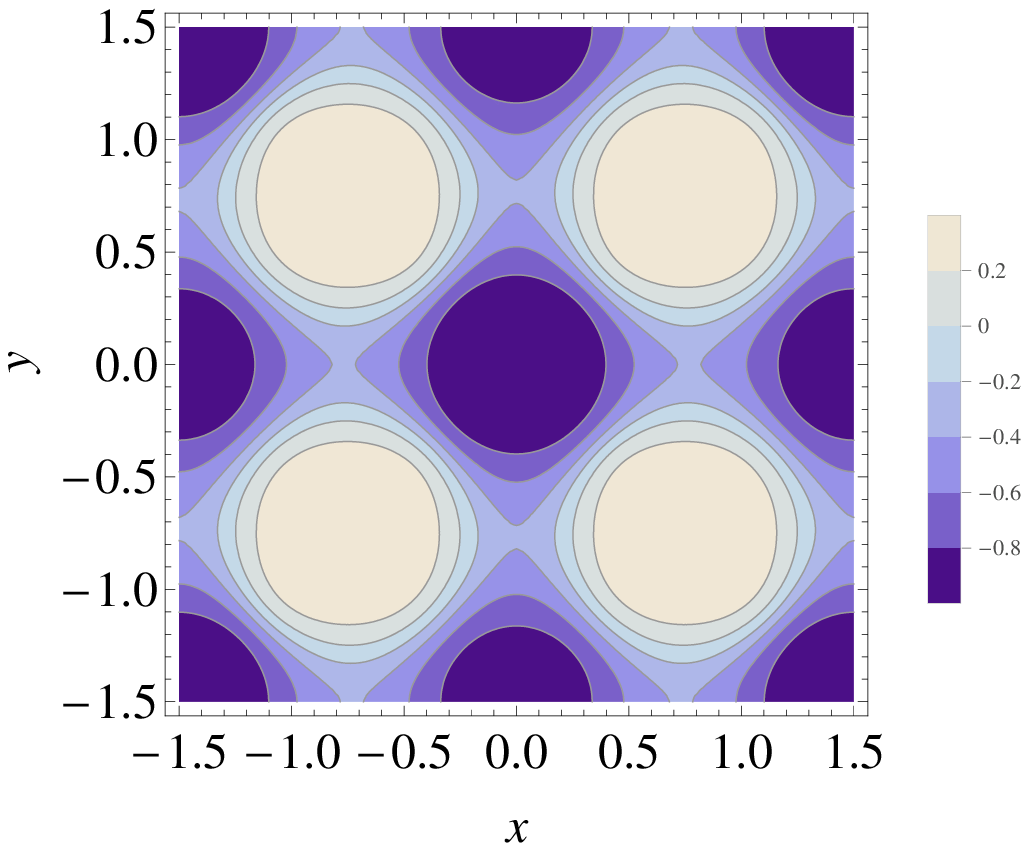} \\
\includegraphics[width=1\textwidth]{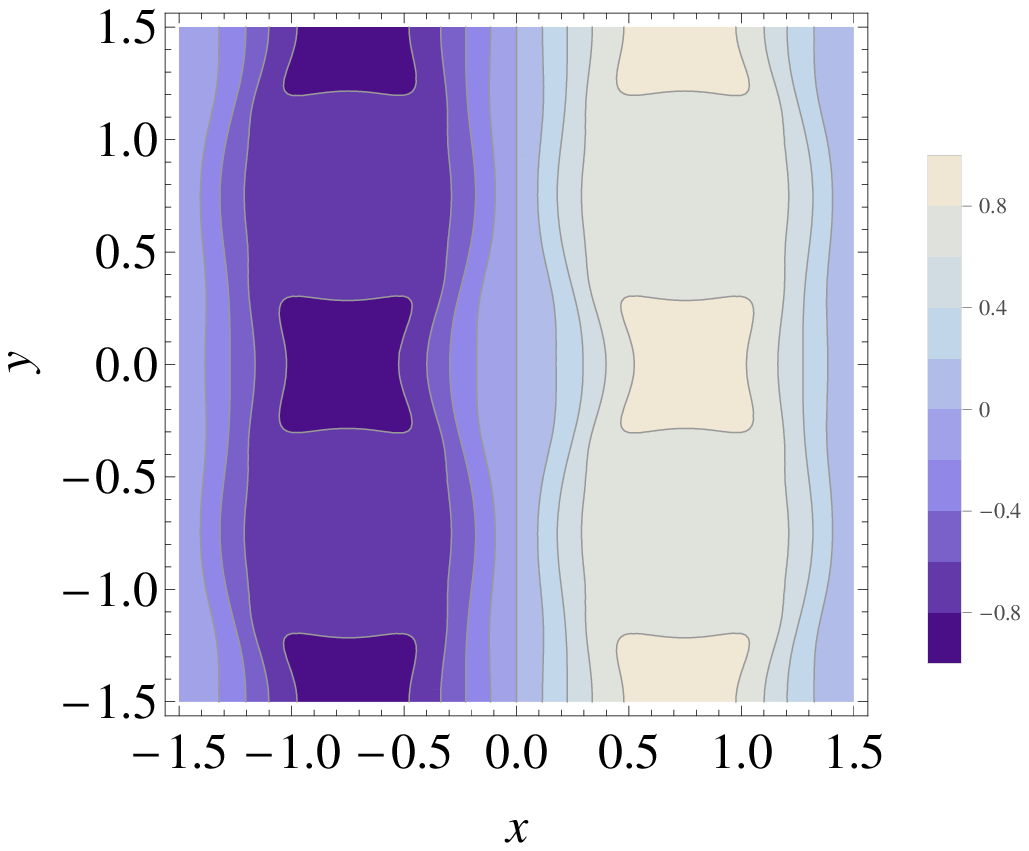}
\end{minipage}
}
\subfigure[~$L=1.4~$fm]{
\begin{minipage}[b]{0.22\textwidth}
\includegraphics[width=1\textwidth]{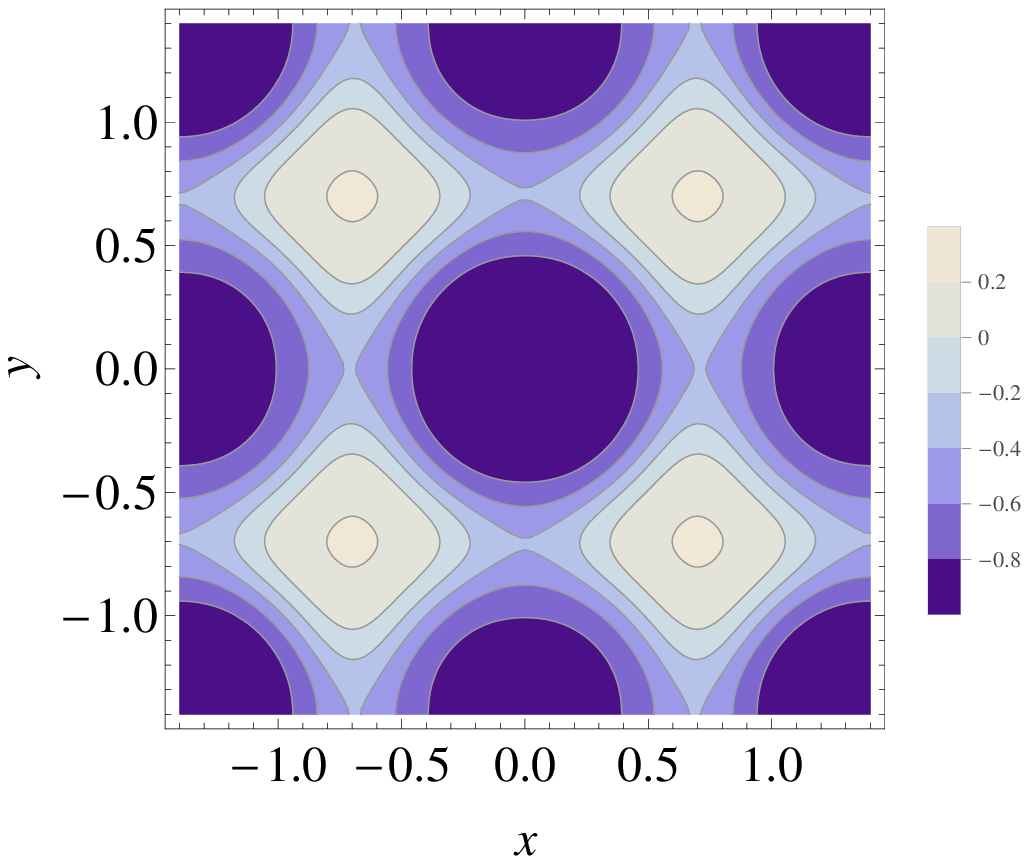} \\
\includegraphics[width=1\textwidth]{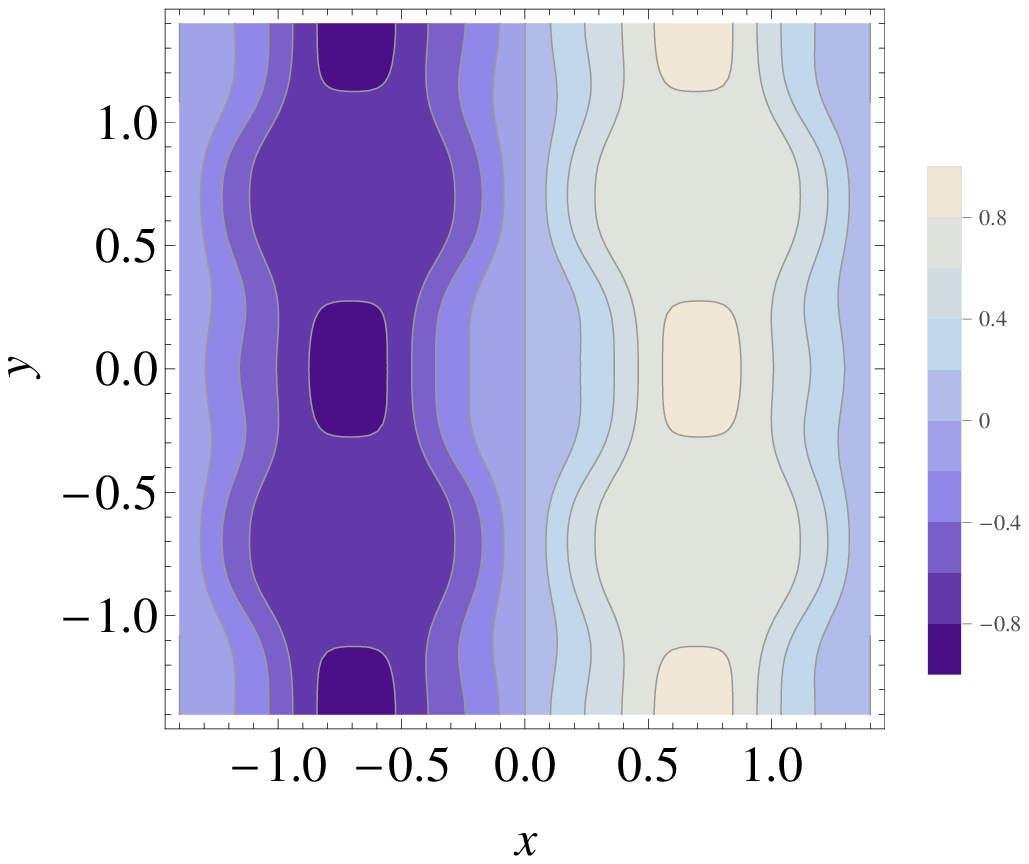}
\end{minipage}
}
\subfigure[~$L=0.9~$fm]{
\begin{minipage}[b]{0.22\textwidth}
\includegraphics[width=1\textwidth]{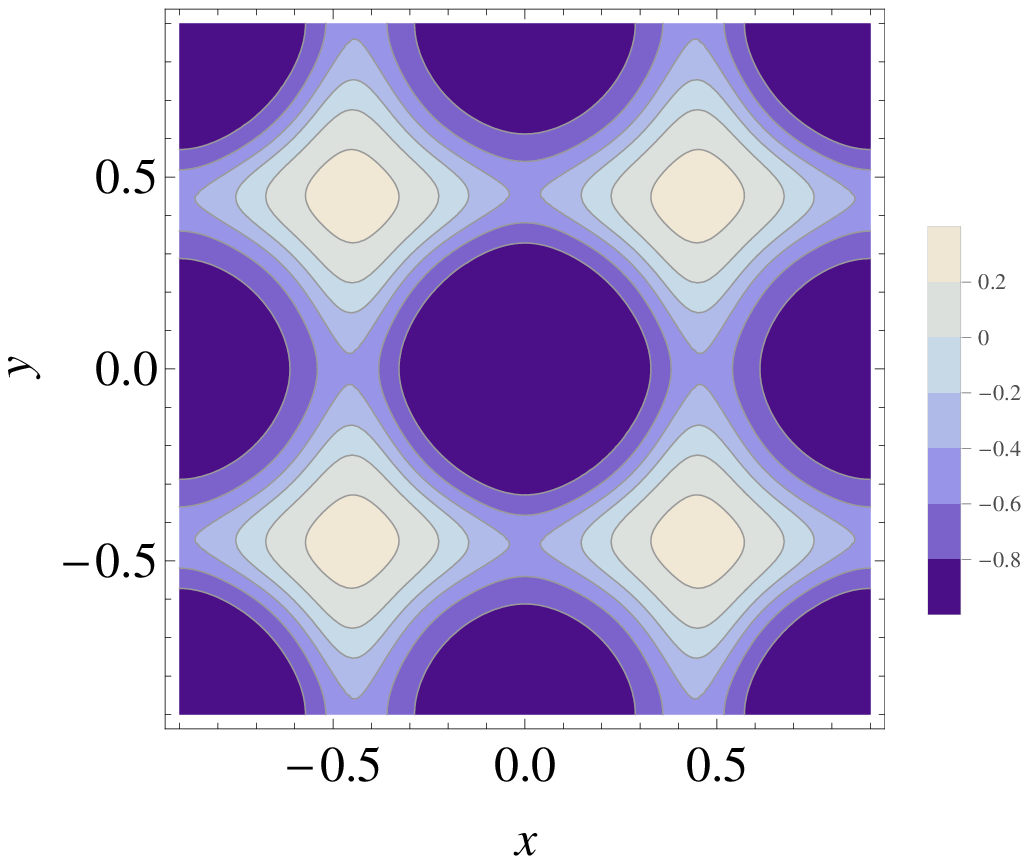} \\
\includegraphics[width=1\textwidth]{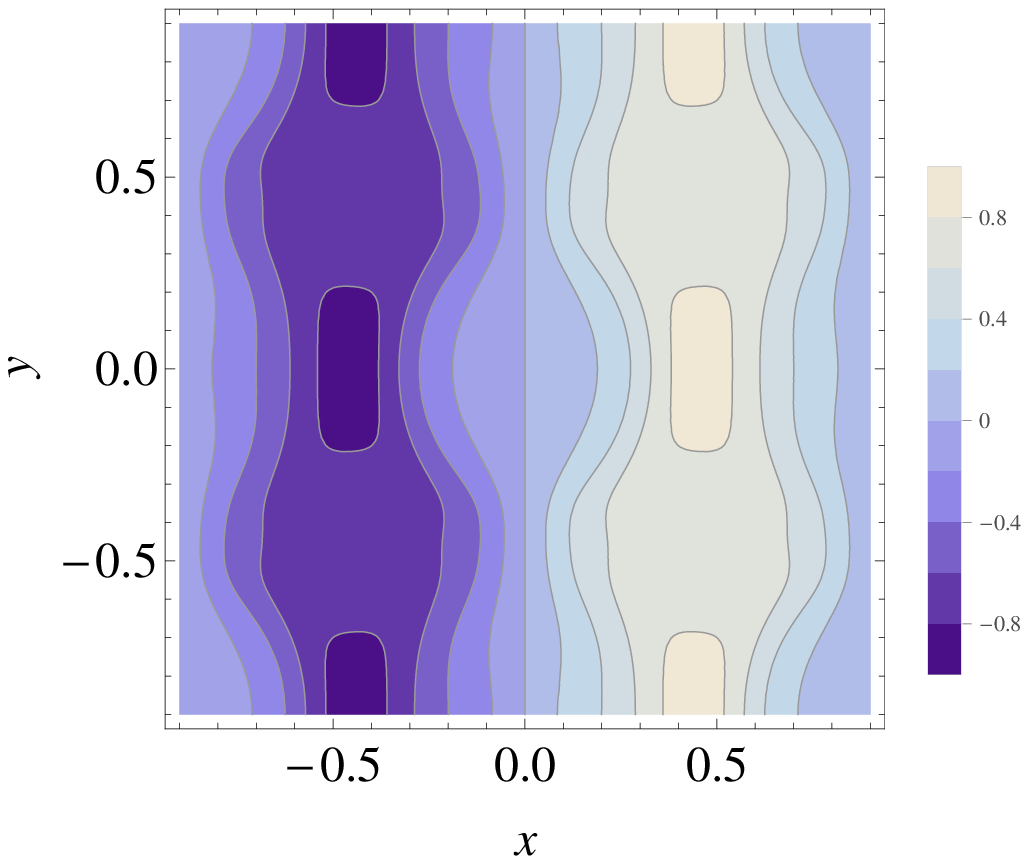}
\end{minipage}
}
\caption[]{(Color online) The density effect on $\phi_0(x,y,0)$ (first row) and $\phi_1(x,y,0)$ (second
row). The half-skyrmion phase appears at $L = 1.45~$fm.}
\label{fig:xydepen}
\end{figure*}

From Eq.~\eqref{eq:norm} we see that, in the skyrmion crystal approach, all the quark condensates are
functions of the three-dimensional coordinates. In Fig.~\ref{fig:xydepen} we illustrate the crystal
size dependence of the quark-antiquark condensates $\phi_0(\vec{x}) \propto \langle \bar{q}q(\vec{x})
\rangle^*$ and $\phi_1(\vec{x}) \propto\langle \bar{q}\tau^1 \gamma_5 q(\vec{x}) \rangle^*$ at the $z =
0$ plane using the model HLS($\pi,\rho,\omega$) in which all the $\pi$, $\rho$, and $\omega$ meson
contributions are included {and a self-consistent power-counting mechanism exists}. These plots tell
us that in both the skyrmion and half-skyrmion phases, the chiral symmetry is locally broken since the
quark condensates could be locally nonzero.

%%%%%%%%%%%%%%%%%%%%%%%%%%%%%%%%%%%%%%%%%%%%%%%%%%%%%%%
%%%%%%%%%%%%%%%%%%%%%%%%%%%%%%%%%%%%%%%%%%%%%%%%%%%%%%%%%

Let us first look at the first row of Fig.~\ref{fig:xydepen}. From Fig.~\ref{fig:xydepen}(a) one sees
that the large area is occupied by white space so that at this density the space average of $\phi_0$
is approximately its value in the QCD vacuum, i.e., $\sim 1$, which agrees with the result of
Ref.~\cite{Ma:2013ooa}. By comparing Figs.~\ref{fig:xydepen}(a) and~\ref{fig:xydepen}(b), we see
that a larger space is occupied by negative values of $\phi_0$, indicating that the space average of
$\phi_0$ becomes smaller as density increases. The comparison of Fig.~\ref{fig:xydepen}(b) -- which
is in the skyrmion phase -- and Fig.~\ref{fig:xydepen}(c) --  which is in the half-skyrmion phase --
indicates that the changeover from the skyrmion phase to the half-skyrmion phase could be smooth.
Finally, the close resemblance of Figs.~\ref{fig:xydepen}(c) and ~\ref{fig:xydepen}(d) reflects the fact that even in
the half-skyrmion phase the chiral symmetry is still locally broken, though the condensate is zero
averaged in the unit cell and the order parameter for chiral symmetry breaking in medium $f_\pi^\ast$
is nearly constant, independent of density.

Next we consider the second row of Fig.~\ref{fig:xydepen} which describes the density dependence of
the inhomogeneous parity-odd isospin nonsinglet quark condensate $\langle \bar{q}\tau^1 \gamma_5 q
\rangle^*$. We first find that this inhomogeneous condensate is an odd function of the coordinate $x$
because in the skyrmion crystal approach the flipping of the coordinate $x_i$ is accompanied by the
flipping of the corresponding component of $\phi_i$ and, because of this property, we conclude that the
space average of the quantity $\phi_i$ is zero. The figures in this row also show the density
dependence of the component $\phi_1$ in the skyrmion phase and its nearly density independence in the
half-skyrmion phase.

\begin{figure*}
\centering
\subfigure[HLS($\pi$)]{
\begin{minipage}[b]{0.22\textwidth}
\includegraphics[width=1\textwidth]{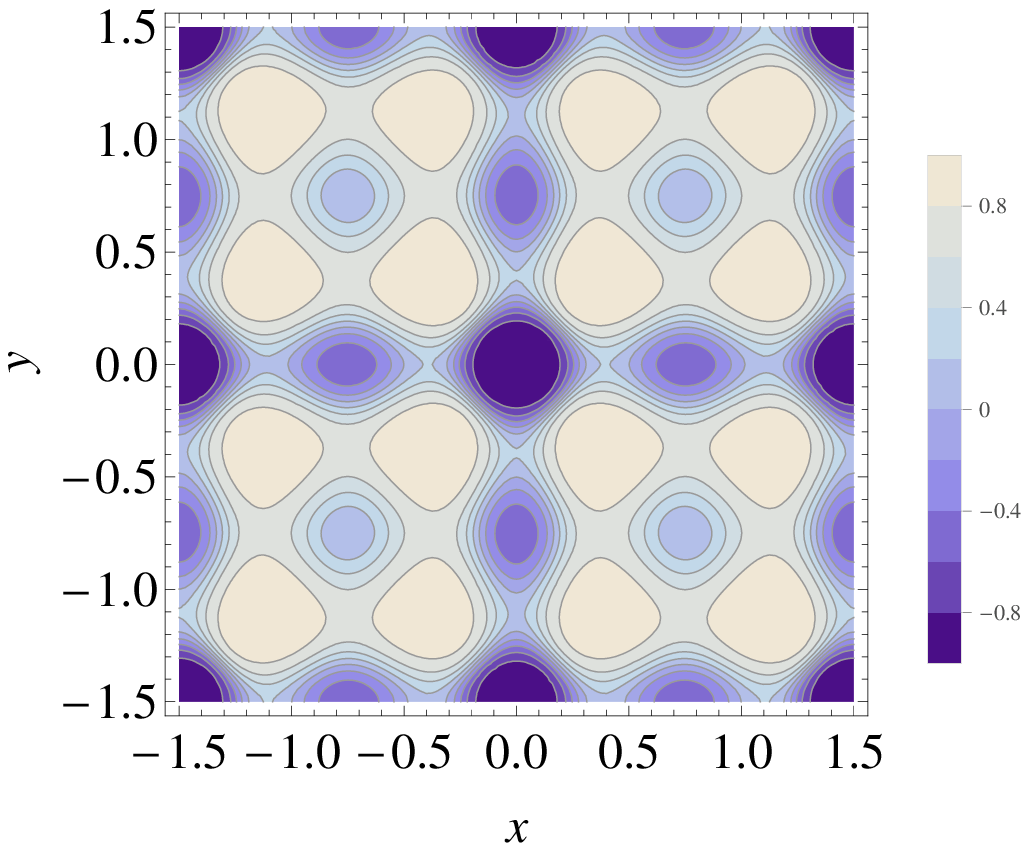} \\
\includegraphics[width=1\textwidth]{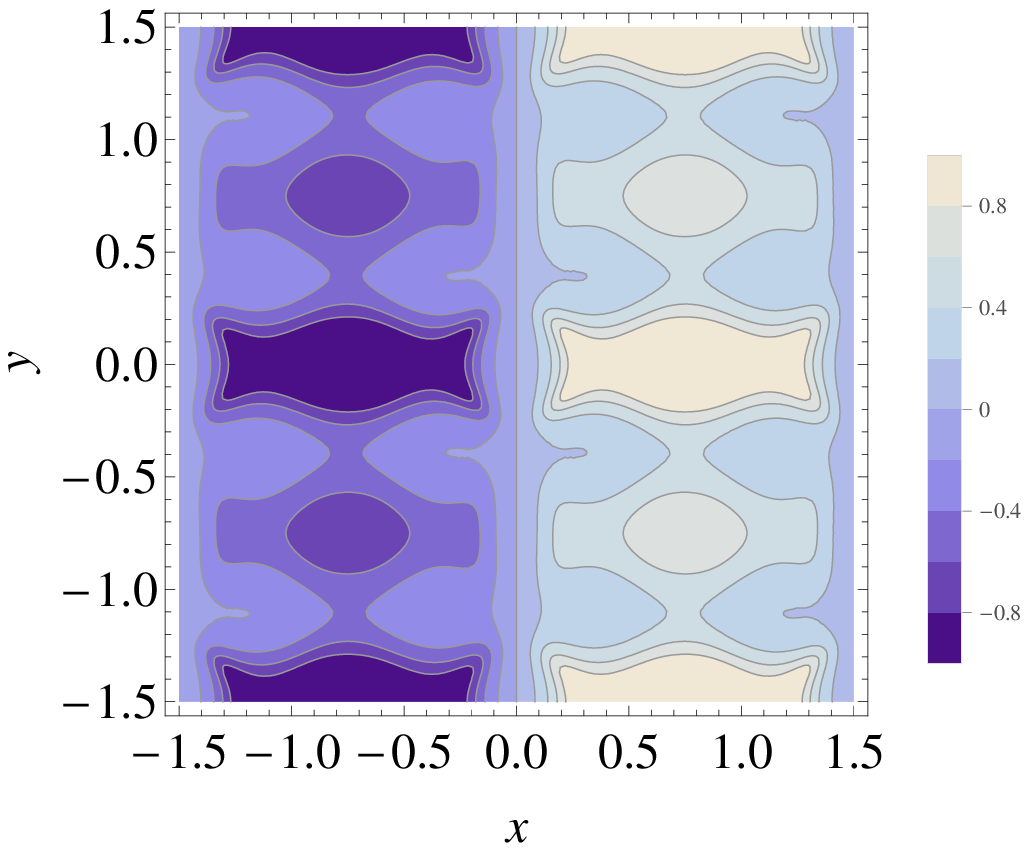}
\end{minipage}
}
\subfigure[HLS($\pi,\rho$)]{
\begin{minipage}[b]{0.22\textwidth}
\includegraphics[width=1\textwidth]{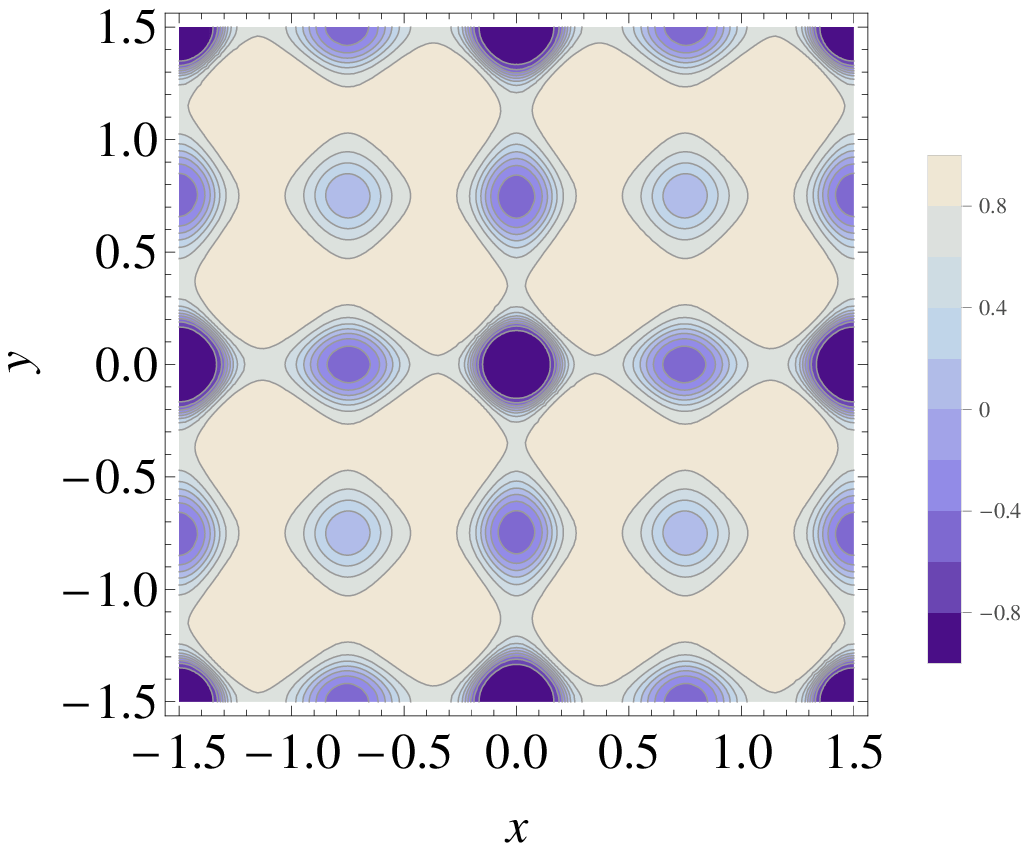} \\
\includegraphics[width=1\textwidth]{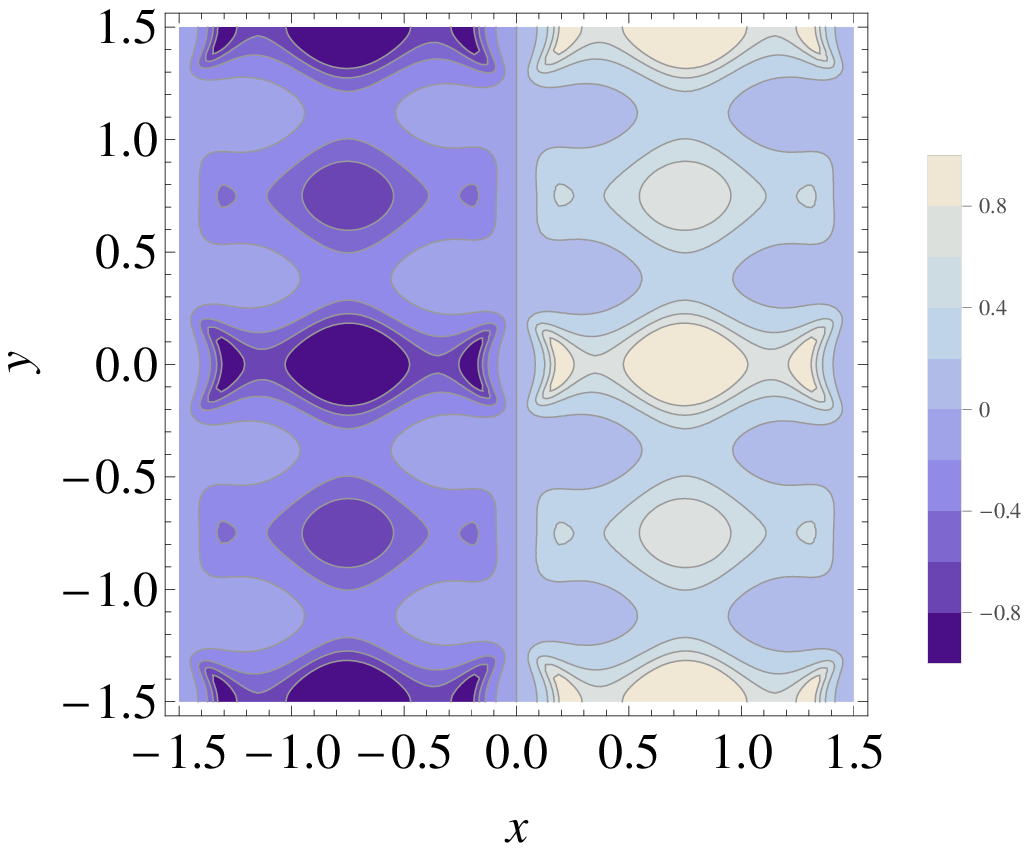}
\end{minipage}
}
\subfigure[HLS($\pi,\rho,\omega$)]{
\begin{minipage}[b]{0.22\textwidth}
\includegraphics[width=1\textwidth]{PlotSigmaprw150Nxy.eps} \\
\includegraphics[width=1\textwidth]{PlotPhi1prw150Nxy.eps}
\end{minipage}
}
\subfigure[dHLS-II($\pi,\rho,\omega,\chi$)]{
\begin{minipage}[b]{0.22\textwidth}
\includegraphics[width=1\textwidth]{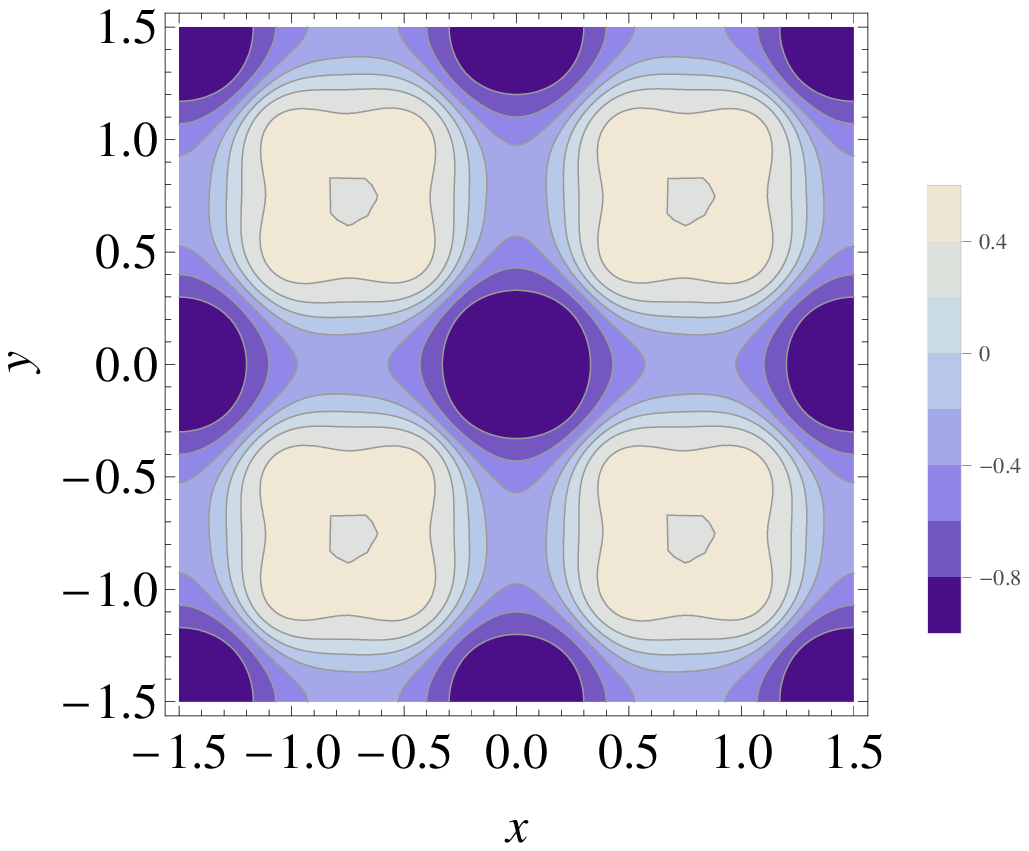} \\
\includegraphics[width=1\textwidth]{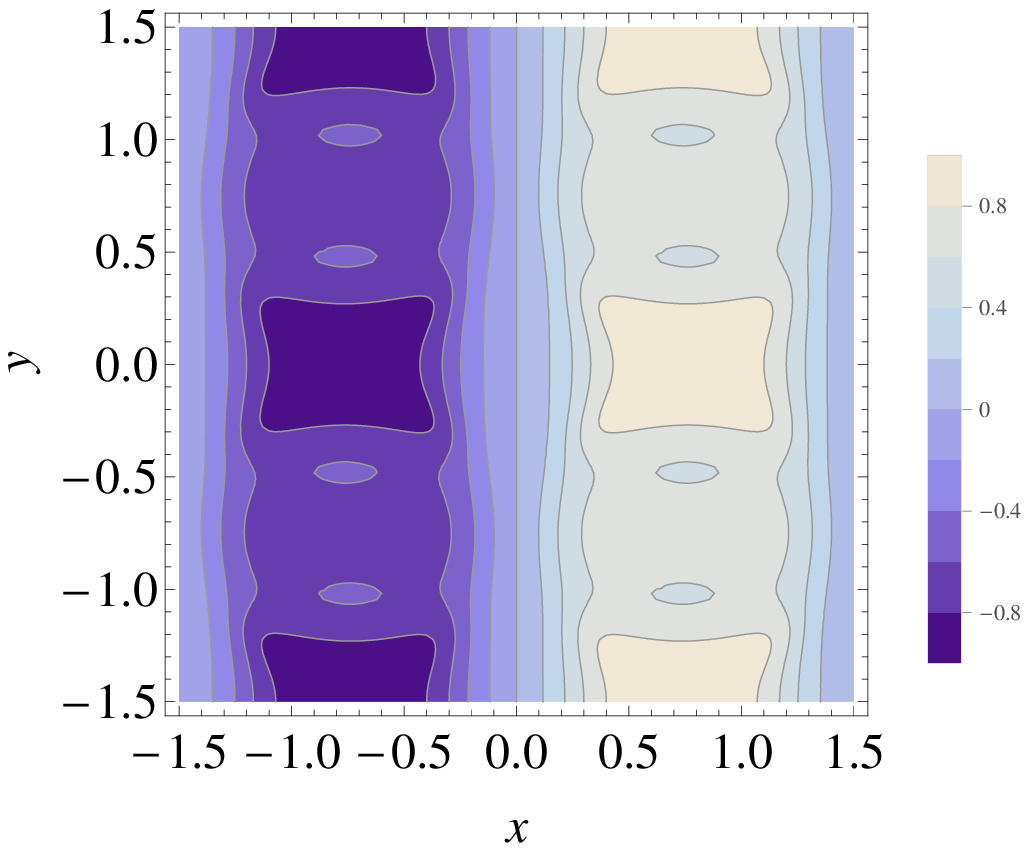}
\end{minipage}
}
\caption[]{(Color online) The $\rho$, $\omega$ and $\chi$ meson effects on $\phi_0(x,y,0)$ (first row)
and $\phi_1(x,y,0)$ (second row) in the skyrmion phase at $L=1.5~$fm.}
\label{fig:mesoneffectxy}
\end{figure*}

We plot in Fig.~\ref{fig:mesoneffectxy} the effect of the $\omega$ meson and the scalar (dilaton)
meson on the quark condensates in the $x$-$y$ plane. In this plot, we take the typical crystal size
$L = 1.5$~fm at which the system is in the skyrmion phase for all four models. Compared to the model
with a pion only, HLS($\pi$), we find that the plot from HLS($\pi,\rho$) is occupied by less blue
(soliton matter) space, which indicates that the quark condensates appear in a smaller area due to the
attractive force from the $\rho$ meson. Similarly, the subtle difference between
Figs.~\ref{fig:mesoneffectxy}(c) and~\ref{fig:mesoneffectxy}(d) tells us that the inclusion of the dilaton field associated
with the dynamical breaking of the scale symmetry slightly shrinks the area of the inhomogeneous
quark condensates due to the attractive force from the dilaton. However, the dramatic difference between
the results in Figs.~\ref{fig:mesoneffectxy}(b) and~\ref{fig:mesoneffectxy}(c) result from HLS($\pi,\rho$) and HLS($\pi,\rho,\omega$)
shows the crucial effect on the quark condensates from the $\omega$ meson. The repulsive force coming
from the $\omega$ exchange causes the inhomogeneous quark condensates to expand. This may explain the
deviation from the BPS structure caused by the presence of the $\omega$ meson noted in item~\ref{here}.

%%%%%%%%%%%%%%%%%%%%%%%%%%%%%%%%%%%%%%%%%%%%%%%%%%%%%%%

As mentioned, in our model, the chiral condensate is locally broken in both the skyrmion phase, and
half-skyrmion phase although in the latter phase it is restored globally. There is a basic difference
between this skyrmion prediction and the CDW phase discussed in Ref.~\cite{Heinz:2013hza}.  First of
all, in~\cite{Heinz:2013hza}, the quark-antiquark condensate is the bona-fide order parameter
indicating the onset of the Wigner phase, whereas it is not in the skyrmion description, where it is more
like a ``pseudo-gap". Second, the ``dilaton" field $\chi$ in \cite{Heinz:2013hza} is what corresponds
to the isoscalar tetraquark state or a pion-pion resonance, whereas our dilaton is $f_0(500)$ which
can be associated with a pseudo-Goldstone boson in the vicinity of a QCD IR fixed point~\cite{CT2015}.
In our scheme, there is no low-lying two-quark configuration fluctuation at low density corresponding
to the fourth component of the chiral four-vector in the linear realization of chiral symmetry.

It is, however, interesting to make a comparison between the two models. To do so, we take the
Dautry-Nyman ansatz~\cite{Dautry:1979bk} as done in  \cite{Heinz:2013hza},
\begin{eqnarray}
\phi_0(x) & = & \cos (2 f x) ; \quad \phi_3 = \sin (2fx).
\label{eq:CDWansatz}
\end{eqnarray}
In this form, the parameter $f$ is an indicator for the inhomogeneous quark condensates. It was found
in \cite{Heinz:2013hza} that, at low chemical potential, $f = 0 $; i.e., there are no inhomogeneous
quark condensates. However, at high chemical potential, $f$ becomes a nearly nonzero constant which
indicates the onset of the inhomogeneous quark condensates.
In terms of the chiral field $U$, the indicator $f$ in \cite{Heinz:2013hza} is simply
\begin{eqnarray}
f & = & - \frac{i}{4}{\rm Tr}\left(\tau_3U_0^\dagger \partial_x U_0 \right).
\label{eq:CDWICond}
\end{eqnarray}
However, in the skyrmion crystal approach, the situation becomes somewhat complicated because of both
the three-space dependence and the three-isospin dependence locked into the hedgehog. Here the relevant
quantity is
\begin{eqnarray}
f_{ij}(x,y,z) & = & - \frac{i}{4}{\rm Tr}\left(\tau_iU_0^\dagger \partial_j U_0 \right)\nonumber\\
& = & \frac{1}{2} \left[\phi_0 \partial_j \phi_i - \phi_i \partial_j \phi_0 + \epsilon_{iab}\phi_a
\partial_j \phi_b\right],
\label{eq:SkyrCond}
\end{eqnarray}
which is a $3 \times 3$ matrix in the spin-isospin space, and each matrix element depends on the
coordinates $x, y$, and $z$. An explicit numerical simulation shows that in the FCC crystal, the
quantities $f_{ij}(x,y,z)$ vanish in neither the skyrmion phase nor the half-skyrmion phase, so the
inhomogeneous quark condensate exists locally in both the skyrmion phase and the half-skyrmion phase. The
inhomogeneity with $f\neq 0$ in the skyrmion phase could be due to the fact that at low density the
classical skyrmion configuration in crystalline form is not in the Fermi liquid state. The homogeneity
may be recovered by collective quantization to restore spin and isospin. The half-skyrmion phase at
$n>n_{1/2}$, however, is likely non-Fermi liquid in a crystalline form.

In the CDW calculation studied in Ref.~\cite{Heinz:2013hza}, the indicator of the homogeneous quark
condensate is the quantity $\phi$ defined there related to the pion decay constant via  $f_\pi =
\phi/Z$, with $Z$ being the wave-function renormalization constant of the pseudoscalar mesons arising from removing the $\pi$-$a_1$ mixing terms. It was found that in the low chemical potential region,
the homogeneous quark condensate $\phi$ decreases slightly with the increase of the chemical potential,
while in the high chemical potential region, it remains as a constant with about $25\%$ of the vacuum.
However, in the FCC skyrmion crystal approach, in all the models considered in the present work, the
homogeneous quark condensate accounted for by the space-averaged $\sigma$ decreases with increasing
density in the skyrmion phase and vanishes in the half-skyrmion phase~\cite{Ma:2013ooa,Ma:2013ela}.

Although the inhomogeneity in the quark-antiquark condensate persists in both the skyrmion phase and
the half-skymrion phase with no obvious phase transitions, there is a drastic change in nuclear
physics across the density $n_{1/2}$~\cite{MR-geb}. As mentioned, parity doubling seems to emerge in the hadronic spectrum suggested in the parity-doubling effective Lagrangian
model~\cite{detar} and lattice studies~\cite{glozman}. Indeed,  it was shown very recently in Ref.~\cite{Suenaga:2014sga}  by coupling the heavy-light mesons with chiral-partner structure to the leading-derivative order HLS($\pi$), that the masses of the chiral partners of the  heavy-light mesons, which are split in the skyrmion matter, become degenerate in the half-skymion phase. It would be interesting to study similar
phenomena in the light meson sector.

%%%%%%%%%%%%%%%%%%%%%%%%%%%%%%%%%%%%%%%%%%%%%%%%%%%%%%
%\acknowledgments

M.~H. was supported in part by the JSPS Grant-in-Aid for Scientific Research (S) No.~22224003 and (c)
No.~24540266.
H.~K.~L. was
supported in part by the WCU project of the Korean Ministry of Education,
Science and Technology (Grant No. R33-2008-000-
10087-0).
The work of Y.-L.~M. was supported in part by the National Science Foundation of China (NSFC) under
Grant No.~11475071 and the Seeds Funding of Jilin University.
%%%%%%%%%%%%%%%%%%%%%%%%%%%%%%%%%%%%%%%%%%%%%%%%%%%%%%%%%%%%%%%%%%%%%%%%%%%%%%%%%%

\end{document}